\documentstyle[epsf]{mn}

\def\Bj{\hbox{$B_J$}} 
\def\Bt{\hbox{$B_T$}} 

\def\Mb{\hbox{{\rm M}$_B$}}
\def\Mpc{{\rm\thinspace Mpc}}
\def\asec{{\rm\thinspace arcsec}}

\def\kms{\hbox{$\km\s^{-1}\,$}}
\def\km{{\rm\thinspace km}}

\def\kpc{{\rm\thinspace kpc}}
\def\magsqsec{\hbox{$\mag\asec^{-2}\,$}}
\def\mag{{\rm\thinspace mag}}

\def\nm{{\rm\thinspace nm}}

\def\refs{\par \noindent \hang}

\def\s{{\rm\thinspace s}}

\def\one_wide{8.5cm}
\def\two_wide{14.0cm}

\title[Fornax and the Seven Dwarfs]
{High surface brightness dwarf galaxies in the Fornax cluster}
\author[M.J. Drinkwater \& M. D. Gregg]
{Michael J. Drinkwater$^1$, and Michael D. Gregg$^2$\\
$^1$School of Physics, University of New South Wales, Sydney 2052, Australia\\
$^2$University of California, Davis, and Institute for Geophysics
and Planetary Physics, \\ Lawrence Livermore National Laboratory,
L-413, Livermore, CA 94550, USA}
\date{To appear in  Mon.Not.R.Astron.Soc.}
 
\begin{document}
 
\maketitle


\begin{abstract} We describe a search for compact dwarf galaxies
in the Fornax cluster using the FLAIR spectrograph on the UK Schmidt
Telescope. We measured radial velocities of 453 compact galaxies
brighter than $B_T\approx17.3$ and found seven new compact dwarf cluster
members that were not classified in previous surveys as members of the
cluster. These are amongst the most compact, high surface brightness
dwarf galaxies known.

The inclusion of these galaxies in the cluster does not change the
total luminosity function significantly but they are important because
of their extreme nature; one in particular appears to be a high
(normal) surface brightness dwarf spiral. Three of the new dwarfs have
strong emission lines and appear to be blue compact dwarfs (BCDs),
doubling the number of confirmed BCDs in the cluster.  We also determined
that none of the compact dwarf elliptical (M32-like) candidates are in
the cluster, down to an absolute magnitude $\Mb=-13.2$. We have
investigated the claim of Irwin et al.\ (1990) that there is no strong
relation between surface brightness and magnitude for the cluster
members and find some support for this for the brighter galaxies
($B_T<17.3$) but fainter galaxies still need to be measured.
\end{abstract}

\section{Introduction}

The determination of galaxy luminosity functions (LFs) is subject to a
number of selection effects.  This is particularly true for the faint
end of the LF in clusters where cluster membership is often
necessarily assigned without recourse to redshifts.  One of the
parameters relied upon to guide visual assignments of membership is
surface brightness (SB).  Disney (1976; see also Davies et al.\ 1988)
argues that most magnitude-limited galaxy surveys are sensitive
to a limited range of surface brightness: high SB galaxies being
unresolved and rejected as `stars' and low SB galaxies falling below
the detection threshold even though their total magnitude is brighter
than the nominal limit. Low SB galaxies have been
studied in detail by Phillipps et al.\ (1987) and Davies et al.\
(1988); in this paper we concentrate on the high SB galaxies.

Normal galaxies with relatively high SB are typically $3$
magnitudes brighter than detection limits in photographic sky surveys
and there is rarely any problem resolving them and distinguishing them
from stars.  In some surveys, however, a selection bias against the
very highest SB compact objects can still occur.  We have been
conducting a spectroscopic survey of the Fornax cluster using the UK
Schmidt Telescope (UKST) FLAIR multi-object spectrograph.  Our aim is
to establish the true cluster luminosity function over as great a
range in luminosity and galaxy types as possible.  It is crucial to
understand any SB selection effects and to look for compact dwarf
galaxies (Holman et al.\ 1995, Drinkwater et al.\ 1997) in particular.
Our starting point for the spectroscopic survey is the Fornax Cluster
Catalog (FCC) of Ferguson (1989) which lists 2678 galaxies in an
area of 40 deg$^2$. The galaxies are classified
as `members' or `background' according to a number of visual
criteria, including morphology and SB, on large scale,
high quality DuPont 2.5m Telescope plates.  Broadly speaking, galaxies
of faint apparent magnitude are classified as members if they have low
SB and background if they have high SB.  This type of visual
classification was used successfully by Binggeli \& Sandage (1984) in
their Virgo Cluster Catalog; in both clusters it works well because
there are voids behind the clusters.

Drinkwater et al.\ (1996) tested the classifications in the Virgo
cluster using spectroscopy and found none of the 300 galaxies measured
were incorrectly assigned.  In the present paper we present results
from a similar, more extensive survey of the Fornax cluster, having
measured 453 galaxy redshifts with FLAIR to determine membership. The
galaxy sample and observations are described in
Section~\ref{sec_samp}: we found seven new compact dwarfs in the
cluster. Their properties are presented in
Section~\ref{sec_prop} and in Section~\ref{sec_disc} we discuss the
implications of our results.  Throughout this paper we use a distance
to the Fornax cluster of 15.4 \Mpc\ and a distance modulus of 30.9 mag
(Bureau, Mould \& Staveley-Smith 1996).


\begin{figure}
\epsfxsize=\one_wide \epsffile{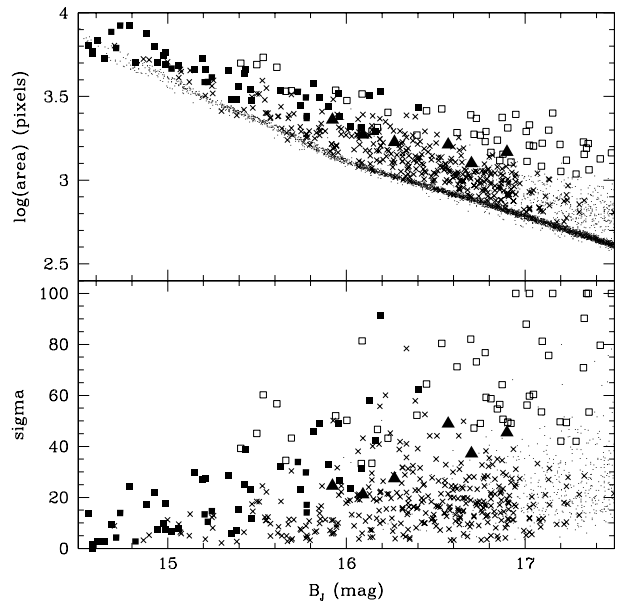}
\centering
\caption{
Image classification diagrams of galaxies in the Fornax
Cluster Catalog. Each point gives the parameters of an image on the
UKST \Bj\ sky survey plate measured by the Automated Plate Measuring (APM)
facility (see Drinkwater et al.\ 1996). The upper and lower
panels respectively plot image area and the classification parameter
`sigma' against apparent magnitude. In both cases the smaller the
parameter at a given magnitude, the more stellar in appearance the
image. Stars are also plotted in the upper panel: these form a
well-defined locus of images with minimum area. The `sigma'
parameter simply measures how many standard deviations a given image
lies above the stellar locus. The FCC-classified members (confirmed:
filled squares; unconfirmed: open squares) are less compact (lower SB)
than the background galaxies (confirmed: crosses; unconfirmed:
dots). The new compact dwarf cluster members are plotted as filled
triangles.
}
\label{fig_memb1}
\end{figure}

\section{Galaxies Observed} 
\label{sec_samp}

Our motivation was to search for any compact cluster members which had
previously been misidentified as background galaxies.  The possible
bias against compact cluster members in the FCC is demonstrated in
Fig.~\ref{fig_memb1} which shows that the FCC-classified `members'
are much less compact than the `non-members' at a given magnitude.
This corresponds to classifying the high surface brightness, faint
apparent magnitude galaxies as background giants. The galaxies we
observed were mostly chosen from the `non-members' with small image
areas in Fig.~\ref{fig_memb1}. The magnitude limit was \Bt=17.3
corresponding to \Bj=17.0 in our calibration.  This sample was chosen
as the most likely to include overlooked compact cluster members and
also to investigate the fate of faded BCDs as discussed by Drinkwater
et al.\ (1996).

We also included many of the galaxies classified in the FCC as
possible `M32-like,' or compact dwarf elliptical galaxies (cdEs), to
investigate the tidal-stripping hypothesis for their formation (Faber
1973; Nieto \& Prugniel 1987).  Preliminary results of this search
were presented by Holman et al.\ (1995).  We observed 78 of the total
of 131 candidates listed in the FCC, an almost complete subsample to a
magnitude limit of \Bt =17.7.  The mean magnitude of Ferguson's M32
candidates of \Bt=17 corresponds to an absolute magnitude of
$\Mb=-13.9$, considerably fainter than M32 itself ($\Mb=-15.5$
(Sandage \& Tammann, 1987)).

\label{sec_obs}

We observed more than 500 candidate cluster members over five
observing seasons with the FLAIR spectrograph on the UKST. We used the
same observing setup as Drinkwater et al.\ (1996) giving a spectral
resolution of 1.3\nm\ and a typical wavelength coverage of
370--720\nm. We reduced the data with IRAF and measured the galaxy
velocities in the heliocentric reference frame by cross-correlation
with template spectra using the RVSAO package (Kurtz et al.\ 1991). We
estimate the total uncertainty in the velocities to be $\pm100\kms$,
based on comparison of the FLAIR measurements with long-slit data.

We measured reliable redshifts for 453 galaxies. The Fornax cluster
is clearly defined in our velocity data with no galaxies in front of
the cluster (less than 780\kms) and a void behind it between 2630 and
4160\kms. We therefore adopted 3000\kms\ as a maximum velocity for
cluster membership giving 39 members and 414 background galaxies from
our data. We combined these with 96 previously published velocities to
give a total velocity sample of 549 galaxies; 113 members and 436
background. In the total sample the cluster has a mean velocity of
1528\kms\ and a total dispersion of 384\kms. These results are used in
Fig.~\ref{fig_memb1} to indicate the confirmed background galaxies and
members.

\begin{figure*}
\centering
\epsfxsize=\two_wide \epsffile{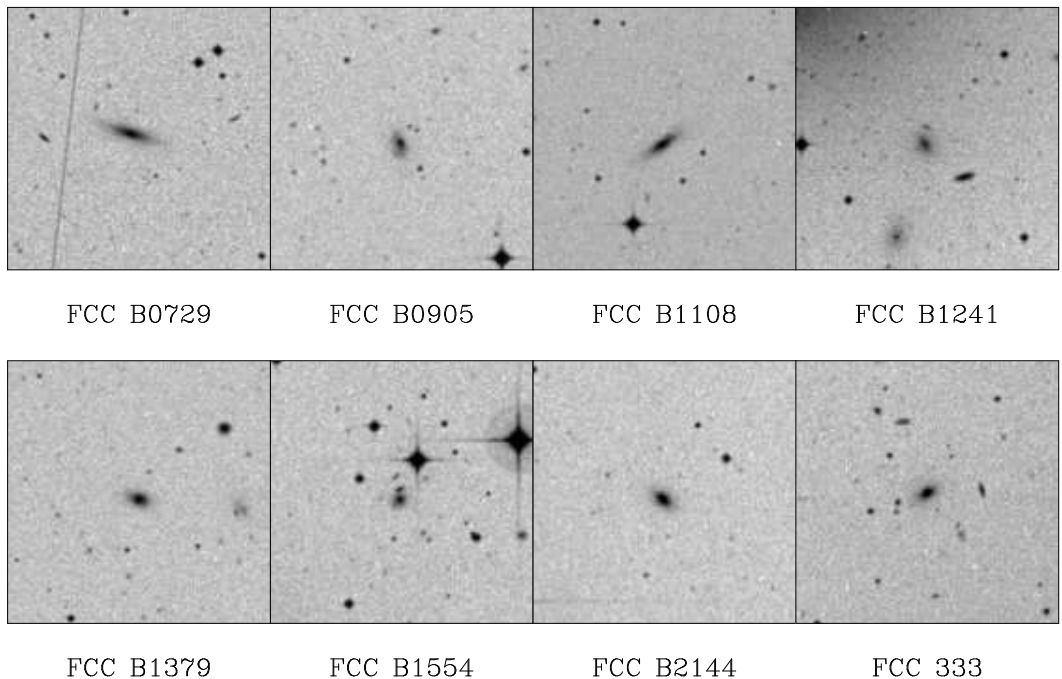}
\centering
\caption{Images of the seven new dwarf galaxies from the Digitized Sky
Survey in the \Bj\ band. The final image is of FCC 333 a galaxy
classified as a member which we have determined is actually a background
galaxy. The
images are all 4 arcmin across with North to the top and East to the
left.}
\label{fig_chart}
\end{figure*}

\begin{figure}
\epsfxsize=\one_wide \epsffile{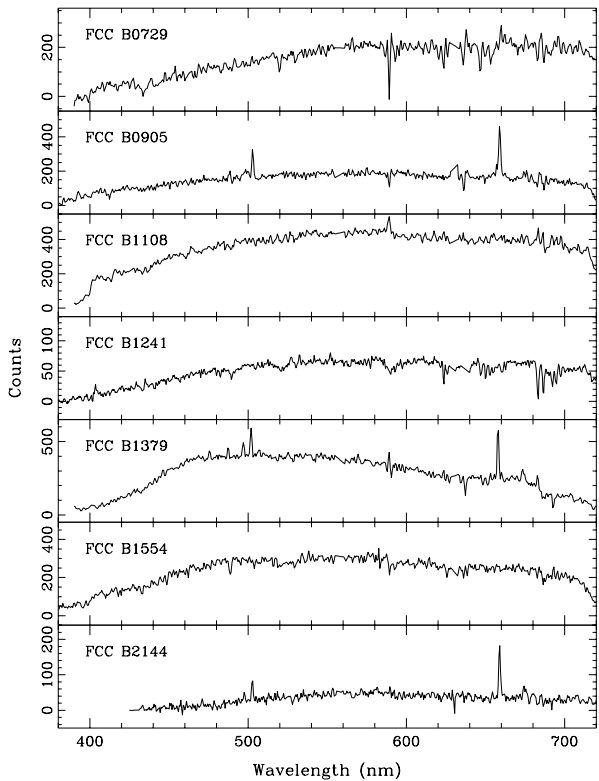}
\centering
\caption{Fibre spectra of the seven new dwarf cluster members. All the
spectra are unfluxed and were taken with FLAIR on the UKST except for 
B1241 which was taken with 2dF on the AAT.}
\label{fig_spec1}
\end{figure}

\begin{figure}
\epsfxsize=\one_wide \epsffile{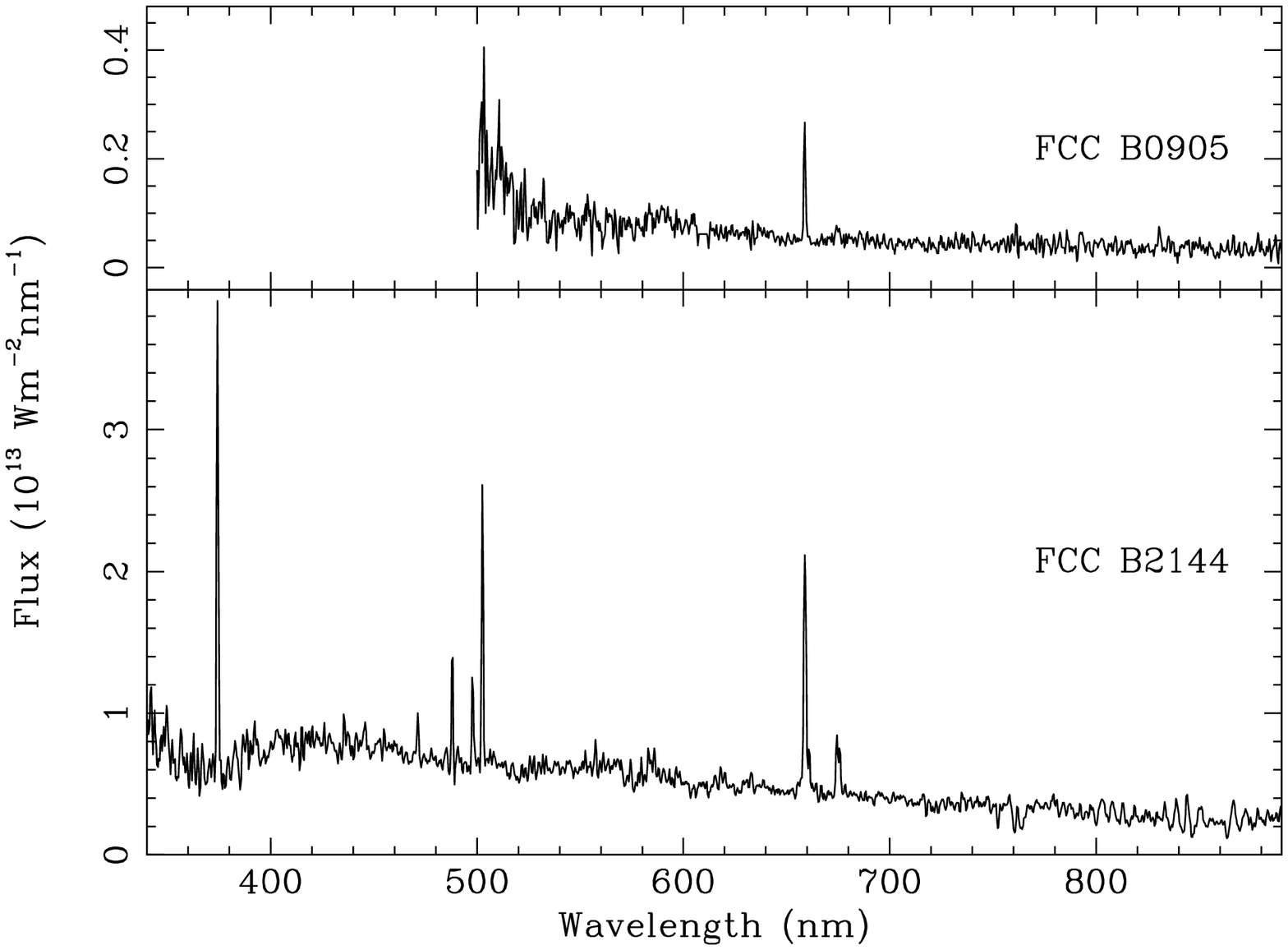}
\centering
\caption{Long-slit spectra of two of the new dwarf cluster members.}
\label{fig_spec2}
\end{figure}

In this paper we concentrate on seven of the newly confirmed members
not classified as likely members in the FCC.  The FLAIR
observations revealed 6 of the new members and one additional cluster
dwarf was identified using the 2dF spectrograph during our first
observations to extend the Fornax redshift survey to fainter limits
using the Anglo-Australian Telescope.  Five of the seven new dwarfs
were classified as `possible members' in the FCC, one as `likely
background' (B0905), and one as `definite background' (B1554).  The
discovery spectra are shown in Fig.~\ref{fig_spec1}; we subsequently
re-observed B1554 with 2dF and obtained higher resolution (8 A)
long-slit spectra of B0905 and B2144 using the Australian National
University (ANU) 2.3m Telescope, all confirming the FLAIR redshifts
(Fig.~\ref{fig_spec2}).  The details of the new dwarfs are given in
Table~\ref{tab_members} and we reproduce photographic \Bj\ images of
them in Fig.~\ref{fig_chart}.  We also include the galaxy FCC 333 for
comparison because it was classified as a definite member, one of the
`M32-type' galaxies listed in the FCC but we found it was a
background galaxy.

\begin{table*} 
\centering
 \begin{minipage}{170mm}

\caption{Details of Observed Dwarf Galaxies}
\begin{tabular}{crllrrrrrl}
      RA (B1950) Dec    &  FCC & M&         Type   & \Bt & \Mb     & SB   &\Bj &  cz (\kms)  & Comment  \\ 
\\      		            		     	      	              	    			                      
3:29:40.00 $-$38:13:54.0& B0729& 3& *dS0 or S0     & 16.5& $-14.4$ &   -- &  --& $1797\pm75$ &          \\ 
3:32:00.40 $-$34:46:38.0& B0905& 4& ?              & 17.0& $-13.9$ &   -- &16.6& $1278\pm50$ &emission  \\ 
3:34:51.70 $-$33:37:24.0& B1108& 3& dS0 or S0      & 16.6& $-14.3$ & 22.2 &15.9& $1941\pm95$ &          \\ 
3:36:21.60 $-$35:40:07.0& B1241& 3& *dE3 pec?      & 17.3& $-13.6$ &   -- &16.9& $1997\pm78$ &          \\ 
3:37:56.80 $-$33:12:43.0& B1379& 3& dE6 pec        & 16.9& $-14.0$ & 22.0 &16.3& $ 708\pm17$ &emission  \\ 
3:40:04.60 $-$35:30:21.0& B1554& 5& E              & 17.5& $-13.4$ &   -- &16.7& $1735\pm50$ &          \\ 
3:47:55.00 $-$32:24:30.0& B2144& 3& E or dE(M32?)  & 16.5& $-14.4$ & 21.8 &16.1& $1155\pm18$ &emission  \\ 
3:48:05.80 $-$36:14:14.0&   333& 1& E or S0        & 16.8& $-14.1$ &   -- &16.4&$13629\pm67$ &background\\ 
\\
\end{tabular}

Notes: FCC, M, Type \Bt\ and \Mb\ are the catalogue number, membership
flag (1=definite member, 3=possible member, 4=likely background,
5=definite background) classification, total magnitude and absolute
magnitude (assuming a distance modulus of 30.9 mag) from the FCC; SB
is the $B$ central surface brightness from our data for B2144 and from
Irwin et al.\ (1990) for the others. The photographic \Bj\ magnitude
is an estimate from our APM data, cz is the heliocentric velocity from
our observations.

\end{minipage}

\label{tab_members}

\end{table*}

\section{Properties of the new dwarf members}
\label{sec_prop}

Fig.~\ref{fig_memb1} shows---as expected---that the new cluster
members are amongst the most compact galaxies in the cluster. This is
also apparent from the images shown in Fig.~\ref{fig_chart}.  Several
have disk-like structure, so it is not surprising they were classified
as giant background galaxies, especially when they are compared to
galaxies like FCC 333 (also shown in Fig.~\ref{fig_chart}) that were
classified as members but we have now shown are background galaxies.

We obtained CCD images of B2144 on the ANU 40-inch Telescope.  An
exponential fit to the $B$ band radial profile has a central SB of
21.8 $B$\magsqsec\ and a scale length of 4.4\asec\ (0.33\kpc). Two of
the galaxies, B1108 and B1379, are in the Irwin et al.\ (1990) sample
with measured SB values of 22.2 and 22.0 $B$\magsqsec\
respectively. These values are typical of compact BCDs (see e.g.
Drinkwater \& Hardy 1991) and confirm the compact nature of these
galaxies.

The fibre spectra of the new dwarfs in Fig.~\ref{fig_spec1} are not
flux-calibrated, but they do show that three (B0905, B1379 and B2144)
have prominent emission lines of H$\beta$, H$\alpha$, and [OIII]~500.7,
indicative of ongoing or recent star formation.  The flux-calibrated
long-slit spectra (Fig.~\ref{fig_spec2}) reveal that both B2144 and
B0905 have blue continua, and B2144 has strong [OII]~372.7, again
pointing to star formation.  We therefore classify both as BCDs and
infer that the other object with emission lines, B1379, is also a BCD.
The absorption line objects are all very weak lined, indicative of
sub-solar abundances, consistent with their dwarf nature.

\section{Discussion}
\label{sec_disc}

\subsection{Compact dwarf ellipticals}

One of the aims of this project was to test the hypothesis
that if compact dwarf ellipticals form through tidal stripping of
ordinary ellipticals, an idea originally due to King (1962), then they
should always be found in association with large companions and should
be rare or non-existent in the outskirts of clusters or as isolated
objects.  We observed 70 of the possible cdEs (Ferguson's Table~XIII
plus additional objects from the background list tagged as `M32?')
in the FCC to see if they occur preferentially in dense environs.
Somewhat surprisingly, we found only a single cluster member in this
subsample; all the rest are background ellipticals.  
The one new cluster member from this sample is B2144, whose spectrum
(Fig.~\ref{fig_spec2}) shows it to be not a cdE but a BCD,
underscoring the uncertainty of classifying objects from imaging data
on morphological grounds alone with no recourse to colours.  The one
cdE with no redshift information classified by Ferguson (1989) as a
`definite member', FCC333, has a velocity of 13629\kms, making it a
background object.

The resulting sample of cdEs has just 7 objects, already tabulated by
Ferguson in the FCC.  Such a small sample makes it difficult to test
the stripping hypothesis, but as 3 are found near large companions and
4 are relatively isolated (Ferguson 1989), tidal stripping appears to
be an unlikely mechanism for the production of cdEs.  This is
supported by the theoretical analysis of Nieto \& Prugniel (1987).

\subsection{Blue compact dwarfs}

In the FCC, a total of 9 candidate BCDs were listed: 4 `definite', one
`likely' and 4 `possible' members according to morphological
classification. Our new observations show that three of these (FCC 24,
B0037 and B0801) are background galaxies and that FCC 35 is a
member. Along with the previously confirmed members FCC 32 and FCC 33,
this gives only 3 BCD members listed in the FCC with spectroscopic
confirmation. The mean absolute magnitude of these is $<$\Mb$> \approx
-16$ (for a distance modulus of 30.9 mag). These BCDs comprise 1\% of
the cluster members, less than the 3\% BCD fraction listed for Virgo
catalogue (Binggeli \& Sandage, 1984). Our observations have detected
an additional 3 new BCDs, effectively doubling the number of confirmed
BCDs although we defer any calculation of the BCD luminosity function
until we can determine an objective (e.g. spectroscopic) definition of
the type. Notably, the new BCDs are significantly fainter than the FCC
BCDs with $<$\Mb$> \approx -14$.

One of the new BCD galaxies, FCC B0905, has a bulge plus disk
morphology and is probably a dwarf spiral (dS) type. Pending future
CCD imaging of the galaxy we estimate its central SB as $\approx23
B\magsqsec$ and scale length as $\approx5\asec$ (0.4\kpc) from our
APM data (used in Fig.~\ref{fig_memb1}). This is the first high
(normal) surface brightness dS to be found. It is distinguished from
spirals in conventional samples by its very small scale length; the 86
spirals measured by de Jong (1996) have scale lengths in the range
1--7\kpc. The discovery of dwarf spirals was previously announced by
Schombert et al.\ (1995) in an HI survey of field galaxies, but these
were much more luminous ($ M_B \approx -16.5$) and more extended (scale
length=1--3\kpc, SB=23--24 $B\magsqsec$) than FCC B0905, and would
have been included in the de Jong (1996) sample.  Schombert et al.\
inferred that dS types are not found in clusters: we suggest rather
that the cluster environment may influence the formation and/or
stability of dS types so that only the smaller, high SB dS types like
FCC B0905 can survive in clusters.

\subsection{Magnitude - surface brightness relation}

Ferguson \& Sandage (1988) found a strong correlation between
surface brightness and magnitude for galaxies they classified as early
type cluster members (SB decreasing with total luminosity). Ferguson
\& Sandage claim that a report of no correlation by Davies et al.\
(1988) was due to contamination by background galaxies.  However
Irwin et al.\ (1990) repeat the claim that the correlation does not
exist but is caused by selection effects in the FCC that cause the
faintest, high SB galaxies and the brightest, low SB galaxies to be
excluded. This disagreement could be resolved with redshifts of the
galaxies to determine unambiguous cluster membership, but the low SB
galaxies are hard to measure.

\begin{figure}
\epsfxsize=\one_wide \epsffile{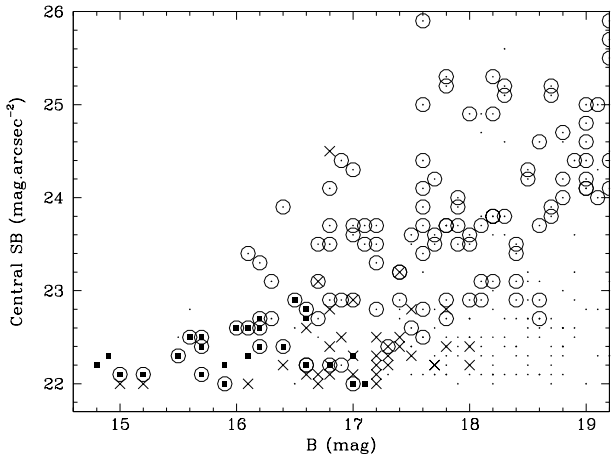}
\centering

\caption{Magnitude - surface brightness diagram of the Irwin et al.\
(1990) low SB galaxy sample with memberships confirmed by our
observations. Filled squares are confirmed members, crosses are
confirmed background galaxies and dots have not been confirmed.
Open circles indicate galaxies classified as dE and dS0 cluster 
members by Ferguson \& Sandage (1988).}

\label{fig_davies}
\end{figure}

In Fig.~\ref{fig_davies} we have replotted the Irwin et al.\
surface brightness and magnitude data for their sample of 321
relatively low SB galaxies in the Fornax cluster.  As a result of our
observations, we now have velocities to confirm 29 of their sample as
members and 37 as background galaxies: these are indicated on the
diagram.  Our data show that at least for magnitudes brighter than
17.3 the cluster members display the full range of SB and no tight
relationship with magnitude. Ferguson \& Sandage (1988) claim their
correlation is only for galaxies they classify as dE and dS0 cluster
members: we have plotted these on the diagram as open circles which do
indeed appear to show a correlation. However the correlation depends
on their having correctly classified all galaxies in the lower
right-hand corner as background giants.  We plan to resolve this issue
finally by measuring these galaxies with 2dF.

\section{Conclusions}

We have confirmed the presence of a number of high (or `normal')
surface brightness dwarf galaxies in the Fornax cluster. These are
among the most compact cluster members and one appears to belong to a
new type of high SB dwarf spiral galaxy, distinct from the dwarf
spirals identified in the field by Schombert et al.\ (1995) which are
2.5 mag more luminous and have SB at least 2 mag fainter. Although these
new dwarfs are quite different to the dS types found by Schombert et
al., the same approach of an HI survey might be used to find more of
them: Barnes et al.\ (1997) found a new, high SB cluster dwarf with
$\Mb=-15.8$ in their HI survey of Fornax.

Additionally, we found no new M32-like (cdE) cluster members in our
survey; all but one of the 78 targets are background objects while the
single member (B2144) is an emission line object with blue colours and
therefore not an M32-like elliptical but a BCD with ongoing star
formation. The distribution of cdEs in the cluster, with 4 of the 7
being relatively isolated, suggests that they are not formed by tidal
stripping by large, nearby companions.

\section*{Acknowledgments}

We particularly thank M. Brown, B. Holman and S. Ryder for making some of the
observations. The 2dF data were used by kind permission of our
collaborators R. Dickens, Q. Parker, S. Phillipps, E. Sadler,
R. Smith and J. Davies (who also provided the data used
in Fig.~\ref{fig_davies}).  Part of this work was done at
the Institute of Geophysics and Planetary Physics, under the auspices
of the U.S. Department of Energy by Lawrence Livermore National
Laboratory under contract No.~W-7405-Eng-48. The NASA/IPAC
Extragalactic Database is operated by the Jet Propulsion Laboratory,
Caltech, under contract with NASA. IRAF is distributed by the
National Optical Astronomy Observatories, operated by the Association
of Universities for Research in Astronomy, Inc.  under cooperative
agreement with the NSF. The Digitized Sky Survey was produced at the
Space Telescope Science Institute under US Government grant NAG
W-2166.

\section*{References}

\refs Barnes, D.G., Staveley-Smith, L., Webster, R.L., Walsh, W., 1997, MNRAS, 288, 307
\refs Binggeli, B., Sandage, A., 1984, AJ, 89, 919
\refs Bureau, M., Mould, J.R., Staveley-Smith, L., 1996, ApJ, 463, 60
\refs Davies, J.I., Phillipps, S., Cawson, M.G.M., Disney, M.J., Kibblewhite, E.J., 1988, MNRAS, 232, 239
\refs de Jong, R.S. 1996, A\&A, 313, 45
\refs Disney, M.J., 1976, Nature, 263, 573
\refs Drinkwater, M.J., Hardy, E., 1991, AJ, 101, 94
\refs Drinkwater, M.J., Currie, M.J., Young, C.K., Hardy, E., Yearsley, J.M.,1996, MNRAS, 279, 595
\refs Drinkwater, M.J., Gregg, M.D., Holman, B.A., 1997 in The Second Stromlo Symposium: The Nature of Elliptical Galaxies, Eds.  M. Arnaboldi, G.S. Da Costa, P. Saha, p. 287
\refs Faber, S.M. 1973, ApJL, 179, 423
\refs Ferguson, H.C. 1989, AJ, 98, 367 (FCC)
\refs Ferguson, H.C., Sandage, A., 1988, AJ, 96, 1520
\refs Holman, B., Drinkwater, M., Gregg, M., 1995, in IAU Colloquium 148: The Future Utilisation of Schmidt Telescopes, eds. J. Chapman, R. Cannon, S. Harrison, B. Hidayat, ASP Conf. Ser. 84, p. 107
\refs Irwin, M.J., Davies, J.I., Disney, M.J., Phillipps, S., 1990, MNRAS, 245, 289
\refs King, I.R. 1962, AJ, 67, 471
\refs Kurtz M.J., Mink D.J., Wyatt W.F., Fabricant D.G., Torres G., Kriss G.A., Tonry J.L., 1991, in  Worrall D.M., Biemesderfer C., Barnes J.,  eds., Astronomical Data Analysis Software and Systems I, ASP Conf. Ser., Vol. 25, p.432
\refs Nieto, J.-L., Prugniel, P. 1987, A\&A, 186, 30
\refs Phillipps, S., Disney, M.J., Kibblewhite, E.J., Cawson, M.G.M., 1987, MNRAS, 229, 505
\refs Sandage, A., Tammann, G.A., 1987, A revised Shapley-Ames catalog of bright galaxies. Carnegie Institution of Washington, Washington, D.C. 
\refs Schombert, J.M., Pildis, R.A., Eder, J.A., Oemler, A, 1995, AJ, 110, 2067

\end{document}